\documentstyle[aas2pp4,epsf]{article}

\received{24 August 1997}
\accepted{4 November 1997}

\slugcomment{Accepted for publication in ApJ}

\lefthead{La Franca, Andreani and Cristiani}
\righthead{Quasar clustering}

\begin{document}

\title{Quasar clustering: evidence for an
increase with redshift and implications for the nature of AGNs}

\author{Fabio La Franca\altaffilmark{1}, Paola Andreani\altaffilmark{2} and
Stefano Cristiani\altaffilmark{2} }

\altaffiltext{1}
{Dipartimento di Fisica, Universit\`a degli studi ``Roma Tre'',
Via della Vasca Navale 84, Roma, I-00146,
{\rm E-mail:} {\tt lafranca@amaldi.fis.uniroma3.it}}
\altaffiltext{2}
{Dipartimento di Astronomia, Universit\`a degli studi di Padova,
Vicolo dell'Osservatorio 5, Padova, I-35122,
{\rm E-mail:} {\tt andreani,cristiani@astrpd.pd.astro.it}}

\begin{abstract}

The evolution of quasar clustering is investigated with a new sample of 388
quasars with $0.3<z\leq 2.2$, $B\leq 20.5$ and
$M_B<-23$, selected over an area of 24.6 deg$^2$ in the South Galactic Pole.
Assuming a two-point correlation function of the form $\xi(r) =
({r/{r_o}})^{-1.8}$, we detect clustering with $r_0=6.2\pm 1.6$ $h^{-1}$
comoving Mpc and $\bar\xi(r= 15~h^{-1} Mpc) = {3\over r^3} \int_0^r
x^2\xi(x)dx =0.52\pm20$ at an average redshift of $<z=1.3>$. We find a
2$\sigma$ significant increase of the quasar clustering between $z=0.95$ and
$z=1.8$, independent of the quasar absolute magnitude and inconsistent with
recent evidence on the evolution of galaxy clustering. If other quasar samples
are added (resulting in a total data-set of 737 quasars) the increase of the
quasar clustering is still favoured although it becomes less significant. With
a parameterization of the evolution of the type $\xi(r,z) =
({r/{r_0}})^{-\gamma}(1+z)^{-(3-\gamma+\epsilon)}$ we find $\epsilon\simeq
-2.5$. Evolutionary parameters $\epsilon > 0.0$ are excluded at a 0.3\%
probability level, to be compared with $\epsilon \sim 0.8$ found for galaxies.
The observed clustering properties appear qualitatively consistent with a
scenario of $\Omega =1$ CDM in which a) the difference between the quasar and
the galaxy clustering can be explained as a difference in the effective bias
and redshift distributions, and b) the quasars, with a lifetime of $t \sim
10^8 yr$, sparsely sample halos of mass greater than $M_{min} \sim
10^{12}-10^{13}~h^{-1}$ M$_{\sun}$. \\ We discuss also the possibility that the
observed change in the quasar clustering is due to an increase in the fraction
of early-type galaxies as quasar hosts at high $z$.

\end{abstract}

\vskip 1 truecm

\centerline{Accepted for publication in the Part 1 of The Astrophysical Journal}
\centerline{Received: 1997 August 24; Accepted: 1997 November 4}

\section{Introduction}

The first detections of the quasar clustering date back more than one decade
(\cite{Shan83}, \cite{Shav84}). Up to now, however, more detailed studies of
the clustering dependence on physical parameters like absolute magnitude and
redshift was hampered by the small number of quasars in statistically
well-defined samples. Recent studies of complete quasars samples
(\cite{Andr92,Mo93,Croo96}) found either no significant difference in the
clustering amplitude measured in comoving coordinates at low ($z<1.4$) and
high ($z>1.4$) redshift, or marginal detections of a decrease of the quasar
clustering with redshift. But the uncertainties were still quite large.

The spatial two-point correlation function, $\xi (r)$, expresses the deviation
of the observed spatial distribution from a random distribution. An estimator
for the correlation function is $\xi(r) = {N^o_p/{N^r_p}} - 1$.
$N^o_p$ is the number of quasars pairs observed at the separations in the range
$r - \delta r/2$ to $r + \delta r/2$, and $N^r_p$ is the number of pairs
counted at the same separation in an artificial random catalogue.
A common parameterization used to describe the evolving correlation function
of galaxies is $\xi(r,z) = ({r/{r_0}})^{-\gamma}(1+z)^{-(3+\epsilon)}$,
where the length is measured in physical (proper) coordinates and $\epsilon$
is an arbitrary fitting parameter. If the separation length is measured in
comoving coordinates the above relation becomes

\begin{equation}
\xi(r,z) = ({r\over{r_0}})^{-\gamma}(1+z)^{-(3-\gamma+\epsilon)}.
\end{equation}

Another estimate of the clustering properties can be obtained via the
integrated pair counts $\Xi (r) = 1 + \bar\xi(r)$, where $\bar\xi (r) =
{3\over r^3} \int_0^r x^2\xi(x)dx$. The advantage in calculating the
integrated two-point correlation function is that it is stable and does not
depend on the binning when $r$ is large. When the boundary effect is
negligible, $\Xi (r)$ is given by $\Xi (r) = {N^o_p(<r)\over{N^r_p(<r)}}$
where $N^o_p(<r)$ and $N^r_p(<r)$ are the number of observed and random
simulated pairs respectively, at a distance less than $r$.
We assume $q_0=0.5$,
$H_0= 100$ Km s$^{-1}$ Mpc$^{-1}$ and comoving distances throughout.

\section{Data}

In order to improve the S/N in the estimate of the clustering evolution, we
have built a new sample of 388 quasars down to $B_J = 20.5$ over a {\it
contiguous} area of 24.6 deg$^2$ with redshift in the range $0.3<z\leq2.2$. 
We define quasars as broad
emission line AGNs having $M_B\leq-23$ with $q_0=0.5$ and $H_0=50$ Km s$^{-1}$
Mpc$^{-1}$. This is the largest sample of quasars of this type yet made. The
candidates were observed with the Meudon-ESO Fibre Optic System (MEFOS,
\cite{Bell}) at the ESO 3.6m telescope.

The sample is centered in the South Galactic Pole region where some of the
Durham/AAT sample areas are included (\cite{Boyl90}), and part of the high
redshift quasar survey of \cite{Warr91} was carried out. 206 quasars are
newly identified. The field is part of the Homogeneous Bright Quasar Survey
(\cite{Cris95}). The quasars were selected through the ultraviolet excess
(UVx) color criterion in the $U-B_j$ versus $B_j-R$ plane. The sample is
divided in two regions: in the central {\it region A} (with an area of 9
deg$^2$), including 176 quasars, the sample is complete (the surface density
is 19.7 quasars/deg$^2$ for $B\leq 20.5$), while in {\it region B} (with an
area of 16 deg$^2$), including 212 quasars, a fraction of the quasar
candidates is still not observed, resulting in an incompleteness of about
$30\%$. The sample with the full quasar catalogue will be described in a
forthcoming paper (La Franca et al., in preparation).

\section{Methods and Results}

Because of the incompleteness in the {\it region B} of the quasar sample, the
overall nonuniform spatial distribution of the quasars complicates
the generation of the "random" data sets used in the computation of the
correlation function. In order to guarantee that the random data set had
exactly the same spatial selection function as the quasar sample, each of the
random data sets was drawn from exactly the same ($\alpha$, $\delta$) positions
as the overall quasar sample. In each random data set, these positions were
assigned different, random, redshifts. We adopted three different methods for
assigning the random redshifts and implemented these in three independent
codes for computing $\bar \xi(r)$. In the three methods the redshifts were
generated (1) random, from the observed $N(z)$ binned in intervals of 0.1 in
$z$; (2) random, according to the distribution expected on the basis of the
quasar Luminosity Function (\cite{Lafr97}); (3) scrambling the observed
redshifts of the quasars in the sample. In the simulation of the quasar
redshifts the sample was splitted into two sub-samples: a bright ($B\leq
18.7$) and a faint ($18.7<B\leq 20.5$) one. This differentiation is necessary
as the bright sample is complete over the whole ({\it A} + {\it B}) area (see
\cite{Cris95}), while the faint sample is the result of several MEFOS
pointings which are complete down to $B=20.5$ only over the central {\it
region A}, but are only locally complete over the remaining {\it region B}.
Each random catalogue was generated 1000 times. The results from these
three methods agree to within the nominal uncertainties obtained from Poisson
statistics.

\begin{figure}
\plotone{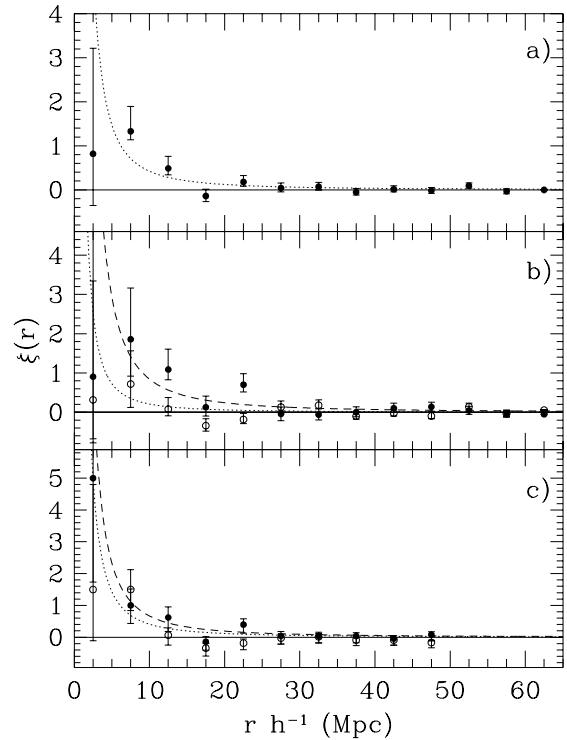}
\caption{ 
a) The autocorrelation function $\xi(r)$ for the total 388
quasars in the SGP sample with $0.3< z\leq 2.2$ as a function of the $r$
comoving distance ($q_0=0.5$, $H_0= 100$). The dotted line is the fit for
$r_0=6.2$ with fixed $\gamma=1.8$. b) The correlation function $\xi(r)$ for the
388 quasars in the SGP sample in two redshift ranges $0.3<z\leq 1.4$, and $1.4<
z\leq 2.2$. The dotted line is the best low-$z$ fit with $r_0=4.2$ while the
dashed line is the best high-$z$ fit with $r_0=9.1$. c) The correlation
function $\xi(r)$ for the complete data-set (737 quasars). The dotted line is
the best low-$z$ fit with $r_0=6.1$ while the dashed line is the best high-$z$
fit with $r_0=8.0$.
}
\label{fig1}
\end{figure}

\placefigure{fig1}

The data set was divided into several luminosity, redshift and spatial
sub-samples in order to study the autocorrelation function $\xi(r)$ and the
integral autocorrelation function $\bar\xi(r)$ as a function of the comoving
distance, and the correlation length $r_0$ assuming a fixed value of
$\gamma=1.8$. The results are summarized in Table 1. In Figure 1{\it a} we
show the autocorrelation function $\xi(r)$ in the range $0.3<z\leq2.2$ for the
total data set of 388 quasars. One sigma Poisson errors in the figure are
based on the number of observed pairs in each bin. The average redshift is
1.34. With the slope fixed to $\gamma = 1.8$ it results $r_0=6.2\pm1.6$
$h^{-1}$ Mpc and $\bar\xi(25,1.34)=0.21\pm0.16$, in agreement with estimate of
Croom and Shanks (1996) of $\bar \xi(25)=0.16\pm0.08$ (their Table 1).

In order to examine the evolution in the amplitude of the correlation function,
the sample was split into the two redshift ranges $0.3<z\leq1.4$, and
$1.4<z\leq2.2$, with average redshift of 0.97 and 1.82 respectively. The
resulting $\xi(r)$ are shown in Figure 1{\it b}. These were fitted by
$\gamma=1.8$ power laws with $r_0$ as a free parameter. At low redshift $r_0=
4.2$ $h^{-1}$ Mpc was found,
corresponding to $\bar \xi(15,0.97) = 0.26\pm 0.27$;
while at high redshift $r_0=9.1$ $h^{-1}$ Mpc, which corresponds to
$\bar\xi(15,1.82)=1.03\pm0.36$, a $1.7\sigma$ significant discrepancy.
The number of observed pairs with $r < 15$ $h^{-1}$ Mpc 
in the range $0.3<z\leq1.4$ is 22 against 17.5 pairs expected.
In the redshift range $1.4<z\leq2.2$, 26 pairs are observed
against 11.3 expected.

In order to check the reliability of this result the same kind of analysis
was applied in the central complete {\it region A} of our data set, using
random samples generated by the three methods described above with the
addition of a random generation of $\alpha$ and $\delta$ coordinates. In the
low redshift subsample, containing 99 quasars, it turns out
$\bar\xi(15,0.95)=0.12\pm0.30$, while in the high redshift subsample,
containing 77 quasars $\bar\xi(15,1.76)=1.38\pm0.48$. Also in this
case the two subsamples show an increase (2.1$\sigma$ significant) of the
clustering amplitude with redshift.

\placetable{tbl-1}

\begin{deluxetable}{l c r c  c c c c}
\tablecaption{Autocorrelation function values with fixed $\gamma =1.8$
\label{tbl-1}}
\tablenum{1}
\tablehead{
\colhead{Area}           & \colhead{Redshift range} &
\colhead{Magnitude range}& \colhead{N$_{QSO}$} &
\colhead{$<M_B>$}          & \colhead{$<z>$} &
\colhead{$\bar\xi(15)$}  & \colhead{$r_0$}  
}  
\startdata
A+B &$0.3<z\leq 1.4$ &      $M_B\leq -23$ &$221$&$-25.0$&$0.97$& $0.26\pm0.27$ & $\phn 4.2\pm \phn 2.5$ \nl
A+B &$0.3<z\leq 2.2$ &      $M_B\leq -23$ &$388$&$-25.6$&$1.34$& $0.52\pm0.20$ & $\phn 6.2\pm \phn 1.6$ \nl
A+B &$1.4<z\leq 2.2$ &      $M_B\leq -23$ &$167$&$-26.5$&$1.82$& $1.03\pm0.36$ & $\phn 9.1\pm \phn 2.0$ \nl
&\cr
A   &$0.3<z\leq 1.4$ &      $M_B\leq -23$ &$\phn 99$&$-24.9$&$0.95$&$0.12\pm0.30$ & $\phn 2.8\pm \phn 4.9$ \nl
A   &$0.3<z\leq 2.2$ &      $M_B\leq -23$ &$176$    &$-25.5$&$1.30$&$0.74\pm0.34$ & $\phn 7.1\pm \phn 1.9$ \nl
A   &$1.4<z\leq 2.2$ &      $M_B\leq -23$ &$\phn 77$&$-26.3$&$1.76$&$1.38\pm0.48$ & $    10.8\pm \phn 2.6$ \nl
&\cr
A+B &$0.3<z\leq 1.4$ & $-27< M_B\leq -25$ &$106$&$-25.7$&$1.11$&$0.09\pm0.31$ & $\phn 2.3\pm 7.5$ \nl
A+B &$0.3<z\leq 2.2$ & $-27< M_B\leq -25$ &$231$&$-26.0$&$1.47$&$0.82\pm0.34$ & $\phn 8.0\pm 2.2$ \nl
A+B &$1.4<z\leq 2.2$ & $-27< M_B\leq -25$ &$125$&$-26.2$&$1.77$&$1.10\pm0.42$ & $\phn 9.5\pm 2.6$ \nl

\enddata
\end{deluxetable}

\placefigure{fig2}
\begin{figure}
\plotone{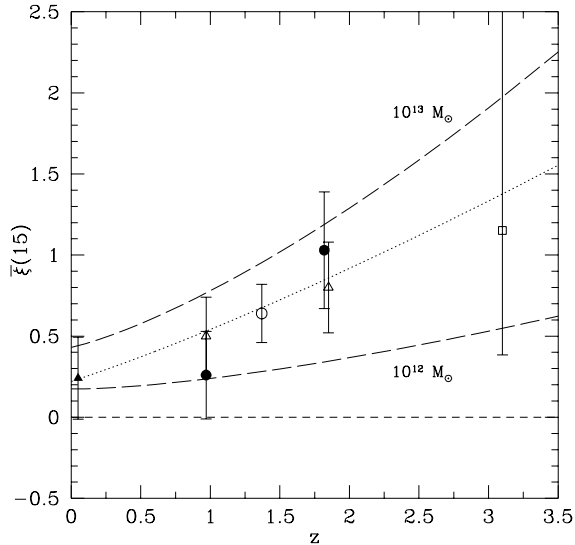}
\caption{ 
The amplitude of the $\bar\xi(15~h^{-1}$ Mpc) as a function of
redshift. The points are the low and high redshift SGP subsamples (filled
circles); our SGP sample plus the high redshift complete quasar samples from
Boyle et al. (1990), La Franca, Cristiani and Barbieri (1992), and Zitelli et
al (1992): complete sample (open circle), and divided in the two redshift
subsamples (open triangles); the low redshift AGN (Seyfert1 and Seyfert2)
amplitude from Boyle and Mo (1993) and Georgantopoulos and Shanks (1994)
(filled triangle); the high redshift sample from Kundi\'{c} 1997 (open
square). The dotted line is the $\epsilon=-2.5$ clustering evolution, while
the dashed lines are the $10^{12}$ and $10^{13}$ $M_{\odot}$ $h^{-1}$ minimum
halo masses clustering evolution according to the transient model of Matarrese
et al. (1997). 
}
\end{figure}

A possible bias in this type of analysis could be introduced by the presence
of large group of quasars. For this reason we have used the {\it friend of
friend} technique to look for groups containing quasars having at least another
quasar at a distance less than 15 $h^{-1}$ Mpc. The technique consists of
drawing a sphere of radius $r$ around each quasar. If there are other friends
within this sphere they are considered to belong to the same group.
The same procedure is applied to the friends until no more friends are
found (see Komberg, Kravtov and Lukash 1996 for details about the method). At
redshift less than 1.4 there are four groups of three members each, while at
redshift higher than 1.4 there are four groups of three members and one
of four members. The four member group is responsible for only 3 of
the total 26 observed pairs at distance less 15 $h^{-1}$ Mpc at high redshift,
while 11.3 are randomly expected. We can, thus, conclude that the result is
not contaminated by the presence in the sample of a rare large group of
quasars at $z>1.4$.

We have also investigated possible biases due to selection effects in the
sample as a function of redshift. Being a flux limited sample, the high
redshift sample selects preferentially brighter quasars. Natural bias
predicts that the amplitude of the autocorrelation function would increase
with galaxy brightness. The high redshift sample has an average
absolute magnitude of $M_B=-26.5$, while the low redshift one has $M_B=-25.0$.
In order to disentangle the magnitude dependence we have computed $\xi$
for the two subsamples with absolute magnitude in the range $-27<M_B\leq-25$.
In this case the average absolute magnitude for the high redshift sample is
$M_B=-26.2$ with $\bar\xi(15,1.77)= 1.10 \pm 0.42$, while for the low redshift
sample is $M_B=-25.7$ with $\bar\xi(15,1.11)=0.09\pm0.31$. Also in this case
the two subsamples show an increase of the clustering amplitude with redshift,
showing that there is no significant dependence of the quasar clustering on the
absolute magnitude.

\section{Comparison with other samples}

Our clustering result has been compared with the clustering computed from
other complete quasar surveys (\cite{Boyl90}, \cite{Lafr92}, \cite{Zite92}).
We excluded the fields included in the SGP from the Durham/AAT survey of Boyle
et al. (1990). A total of 349 quasars have been collected. As already discussed
by Andreani and Cristiani (1992), these samples show no significant evolution
in comoving coordinates.  However, by adding our new sample to these data (a
total data-set of 737 QSOs) the statistical significancy on the increase of
the quasar clustering evolution with redshift is reduced. The resulting
$\xi(r)$ are shown in Figure 1{\it c}. At low redshift we find $\bar
\xi(15,0.97) = 0.50\pm 0.24$ ($r_0=6.1$ $h^{-1}$ Mpc), while at high redshift
$\bar \xi(15,1.85) = 0.80\pm 0.28$ ($r_0=8.0$ $h^{-1}$ Mpc). The total number
of observed pairs with $r < 15$ $h^{-1}$ Mpc in the range $0.3<z\leq1.4$ is 39
against 26 pairs expected. In the redshift range $1.4<z\leq2.2$ a total of 42
pairs are observed against 23.3 expected. In the complete redshift range
$0.3<z\leq2.2$ we find an average value of $\bar \xi(15,1.37) = 0.64\pm 0.18$
corresponding to $r_0=7.0$ $h^{-1}$ Mpc.

In Figure 2 the evolution of $\bar \xi(15)$ as a function of redshift is shown.
We compare our result with previous analyses of the quasar clustering at low
and high redshift. \cite{Boyl93} obtained clustering statistics for 
low redshift ($z<0.2$) quasars in the {\it Einstein} Extended Medium
Sensitivity  Survey (EMSS). Below 10 $h^{-1}$ Mpc four quasar pairs are
observed compared with the prediction of 2.37 pairs ($\bar \xi (10)=0.7)$.
\cite{Geor94} carried out a similar analysis on the low redshift ($z <0.1$)
Seyferts 1 and 2 galaxies in the IRAS Point Source Catalogue finding $\bar \xi
(20)=0.52\pm0.13$. For the Seyfert 1 only sample it turns out $\bar \xi
(20)=-0.10\pm0.27$. In order to compare these results with our data, we added
together the EMSS and IRAS samples and extrapolated the numbers of pairs below
10 and 20 $h^{-1}$ Mpc respectively to a distance of 15 $h^{-1}$ Mpc, assuming
a power law slope $\gamma = 1.8$. This point is shown in Figure 2 with
$\bar\xi(15,0.05) = 0.24\pm0.25$.

We have fitted the two points at low and high redshift (obtained from our
sample plus the quasar samples from Boyle et al. (1990), La Franca, Cristiani
and Barbieri (1992), and Zitelli et al. (1992)) and the point from the EMSS and
IRAS sample with the $\xi$ dependence on redshift expressed in eq. 1, assuming
$\gamma = 1.8$. We find $\epsilon=-2.5\pm1.0$ with
$\bar\xi(15,0)=0.22\pm0.20$, which corresponds to $r_0(z=0)=3.9$ $h^{-1}$ Mpc
or $r_0(z=1.5) = 7.3$ $h^{-1}$. From $\chi^2$ statistics the two values of
$\bar\xi(15)$ in the redshift range $0.3<z\leq2.2$ are
indistinguishable from a constant comoving clustering evolution. However,
statistics are sufficient to constrain the quasar clustering evolution. All the
evolution models normalized to the $\bar\xi(15,0.93)=0.51$ average quasar and
local Seyfert clustering and with $\epsilon >0.0$ are rejected at a
probability level less than 0.003, while the models with a normalization to
the $\bar\xi(15,1.37)=0.64$ average quasar clustering and with $\epsilon >
-0.5$ are rejected at a probability level less than 0.003. As discussed
in the next section, this result should
be compared with $\epsilon \sim 0.8$ found for galaxies.

A recent analysis of the quasar correlation function in the Palomar Transit
Grism Survey has found that $\xi(z>2)/\xi(z<2)=1.8^{+2.5}_{-1.2}$ (Kundi\'{c}
1997), with an average redshift of 1.3 for the low-redshift subsample and 3.1
for the high redshift one. The corresponding high redshift point
($\bar\xi(15,3.1)=1.2$) is plotted in Figure 2 and was obtained by multiplying
our low-redshift $\bar \xi (r=15)$ measurement at $z=1.37$ by
$\xi(z>2)/\xi(z<2)$ given by \cite{Kund97}. Stephens et al. (1997) from an
independent analysis of the same data-set find an even higher value with
$r_0(z=3.3) = 17.5$ $h^{-1}$ corresponding to $\bar\xi(15,3.3)=3.3$. These
analyses, although uncertainties on the clustering of this high redshift
sample are very large, strengthen the evidence for an increase of the quasars
clustering amplitude with redshift and are in agreement with the $\epsilon =
-2.5$ evolution law within a 1$\sigma$ confidence level.

\section{Limits imposed by the XRB ACF}

Studies of the X-ray background (XRB) residual fluctuations and their
autocorrelation function (ACF) have been used to constrain the counts,
evolution and clustering properties of the extragalactic X-ray sources, in
particular quasars. Higher values of $r_0$ and lower values of $\epsilon$
determine a larger contribution of quasars to the soft XRB fluctuations, with
the possible result of exceeding the limits on the XRB ACF and spectral shape.
The general consensus (e.g. \cite{Carr92}; \cite{Dane93}, \cite{Solt94}) is
that a population of sources, clustered like quasars, cannot produce more than
$30 \%$ - $50\%$ of the still unresolved XRB. Soltan and Hasinger 1994 show
(their Figure 8) that if quasars contribute 100\% of the total XRB they would
require a correlation length of 1.5-2 $h^{-1}$ Mpc at $z=1.5$ with
$-1.2<\epsilon<2.4$. If instead a quasar contribution of 35\% is assumed, a
correlation length of 4.5-6 $h^{-1}$ Mpc is allowed.

A recent analysis of deep ROSAT fields (\cite{Mcha97})
has shown that quasars (i.e. broad emission line AGNs) unlikely contribute
to more than 40 per cent (but at least 30\%) of the total XRB at 1 KeV. In the
framework of this result, our outcomes agree with the limits estimated by
\cite{Dane93}. Indeed, from Danese et al. (1993, their Figure 3), it is
possible to see that our estimate of $r_0(z=0) = 3.9$ $h^{-1}$ Mpc and
$\epsilon \sim -2.5$ allows a contribution of more than 15\% for fluxes
fainter than $5 \times 10^{-15}$ erg cm$^{-2}$ s$^{-1}$ in the 0.9-2.4 KeV
band. At these fluxes quasars contribute already at $\sim28\%$ of the total
XRB (see \cite{Mcha97} their Figure 7) and this corresponds to an allowed
contribution of quasars to the total XRB of more than 45\%, in agreement with
the limits imposed by McHardy et al. (1997).

\section{Discussion}

An increase of the clustering amplitude with redshift is a powerful diagnostic
for models of quasar formation and evolution. Schematic patterns of the type
``continuous-activity'', in which each AGN undergoes a long, continuous
dimming after a short formation phase at $z>2$, and ``recurrent-activity'', in
which the potential-AGN population shows intermittent episodes of brightening
lasting a relatively short time, cannot be easily reconciled with the present
observations. If the same population of mass condensations, formed at a given
redshift, is permanently or recurrently shining as quasars, then in any
hierarchical clustering scenario an increase of the clustering with decreasing
redshift would be expected. On the other hand, the quasar phenomenon may
represent a single, short event ($\tau \sim 10^8$ yr) in the host galaxy
lifetime. If we associate it with a characteristic mass at all epochs, i.e. a
sparse sampling of the distribution of halos with mass greater than $M_{min}$,
quasars would correspond to rarer and rarer over-densities with increasing
redshift and their clustering amplitude is expected to grow with $z$. This
scenario is described as the {\it transient model} by \cite{Mata97}.

In a more quantitative way we can follow Matarrese et al. (1997) in estimating
the observed correlation function in a given redshift interval in the
framework of an $\Omega=1$ CDM model. This is carried out by suitably weighting
the mass autocorrelation function $\xi(r,z)$ with the number of objects as a
function of redshift and convolving with the effective bias factor $b_{eff}$.
The effective bias $b_{eff}(z)$ is the weighted mean of the bias $b(M,z)$ with
the mass distribution (e.g. Press-Schecter) at a given redshift. Figure 2
shows that for a minimum halo mass in the range $M_{min} = 10^{12} - 10^{13}
h^{-1} M_{\odot}$, the QSO two-point correlation function is reproduced both
in absolute value and redshift evolution. It is interesting to note that
\cite{Cav97}, on the basis of the shape and evolution of the QSO Luminosity
Function, suggest that the quasar phenomenon at $z < 3$ is connected with
interactions in relatively small galaxy groups, whose typical mass, $5 \cdot 
10^{12}~ M_{\odot}$, is remarkably similar to our estimate, based on the
clustering properties.

What is the relation between the possible observed increase of the quasar
clustering with redshift, the recent evidence of a decrease of the galaxy
clustering with redshift with $\epsilon \sim 0.8$ (e.g. \cite{Lefe96},
\cite{Carl97}, \cite{Vill97}) and the theories of formation and evolution
of cosmic structure?

We can schematically envisage that either galaxies form via the merging of
lower-mass halos or they form at some characteristic redshift (the {\it
merging} and {\it object-conserving} models of Matarrese et al. 1997,
respectively). In this framework the non-linear processes leading to the
galaxy formation give origin to a bias parameter with a distinctive evolution.
The convolution of the bias $b(M,z)$ with the mass autocorrelation function
$\xi(r,z)$ and the galaxy redshift distribution can produce the observed
decrease of the galaxy clustering with increasing redshift.

It must be pointed out, however, that different population of galaxies may
show different clustering properties and there are indications that for
high-$z$ star-forming galaxies (\cite{Steid})\, the clustering amplitude does
not decrease so quickly as for faint low-$z$ galaxies (\cite{Lefe96}).

Another (not necessarily alternative) scenario, which can explain the QSO
clustering behaviour, is based on a change of the morphological mix of
the quasar host galaxies with redshift. Recently \cite{Fran97} have proposed,
on the basis of the metallicities and abundance ratios in absorption-line
systems associated with quasars, that the host galaxies at high redshift are
early-type bulge dominated, formed quickly through a huge episode of star
formation. This picture is reinforced by the observed increase with redshift
of the dust mass associated with quasars (\cite{Andr97}). On the other hand,
low-$z$ quasars may be related to the accretion onto black holes in the nuclei
of late-type galaxies, where gas is still available. At low redshift
early-type galaxies are observed to cluster more strongly than late type
galaxies (\cite{Love95}) by a factor $3-5$. If this behaviour persists at all
redshifts, an increase of the fraction of elliptical galaxies as quasar hosts
could produce the increase of the quasar correlation at high redshifts.

\acknowledgments

FLF acknowledges the hospitality of the {\it Royal Observatory of Edinburgh}.
We acknowledge the use of UKST plates, and of the COSMOS plate scanning
device. We thank Alberto Franceschini, Francesco Lucchin,
Sabino Matarrese, Lauro Moscardini, Cesare Perola, Simon White for useful
discussions, and Claudio Lissandrini, Mike Hawkins, Harvey McGillivray,
Lance Miller who collaborated in the selection of the quasar candidates. This
work was partially supported by the ASI contracts 94-RS-107 and 95-RS-38.

\end{document}